\documentclass[%
 reprint,
superscriptaddress,
 amsmath,amssymb,
 prd,
]{revtex4-2}
\usepackage{xcolor}
\usepackage{graphicx}%
\usepackage{dcolumn}%
\usepackage{bm}%
\usepackage{times}%
\usepackage{hyperref}%
\usepackage[normalem]{ulem}
\hypersetup{
  colorlinks=true,        %
  linkcolor=blue,         %
  citecolor=cyan,         %
  urlcolor=cyan            %
}
\usepackage{CJKutf8}
\usepackage{orcidlink}
\usepackage{soul}

\makeatletter
\newcommand{\refjnl}[1]{{\rmfamily#1}}
\makeatother

\graphicspath{{./}{figures/}}

\begin{document}
\begin{CJK*}{UTF8}{gbsn}

\title{
{Decoupling of a supermassive black hole binary\\ from its magnetically arrested circumbinary accretion disk}
}

\author{Elias R. Most \orcidlink{0000-0002-0491-1210}}
\affiliation{TAPIR, Mailcode 350-17, California Institute of Technology, Pasadena, CA 91125, USA}
\affiliation{Walter Burke Institute for Theoretical Physics, California Institute of Technology, Pasadena, CA 91125, USA}
\author{Hai-Yang Wang(王海洋) \orcidlink{0000-0001-7167-6110}}
\affiliation{TAPIR, Mailcode 350-17, California Institute of Technology, Pasadena, CA 91125, USA}

\begin{abstract}
\noindent 
Merging supermassive black hole (SMBH) binaries will likely be surrounded by a circumbinary accretion disk. Close to merger, gravitational radiation-driven inspiral will happen on timescales faster than the 
effective viscous time at the disk cavity wall, leading to a decoupling of the inner binary dynamics from the surrounding gaseous environment.
Here we perform the first simulation of this decoupling process from a magnetically arrested circumbinary accretion disk. In this regime, the central cavity is filled with very strong vertical magnetic flux, regulating accretion onto the binary. Our simulations identify three main stages of this process: (1) Large-scale magnetic flux loss prior to decoupling. (2) Rayleigh-Taylor-driven accretion streams onto the binary during and after decoupling, which can power magnetic tower-like outflows, resembling dual jets. (3)
Post merger, the cavity wall becomes unstable and the magnetic flux trapped inside the cavity will get ejected in large coherent outbreak episodes with implications for potential multi-messenger transients to merging SMBH binaries.
\end{abstract}

\maketitle
\end{CJK*}

\begin{figure*}
    \centering
    \includegraphics[width=\linewidth]{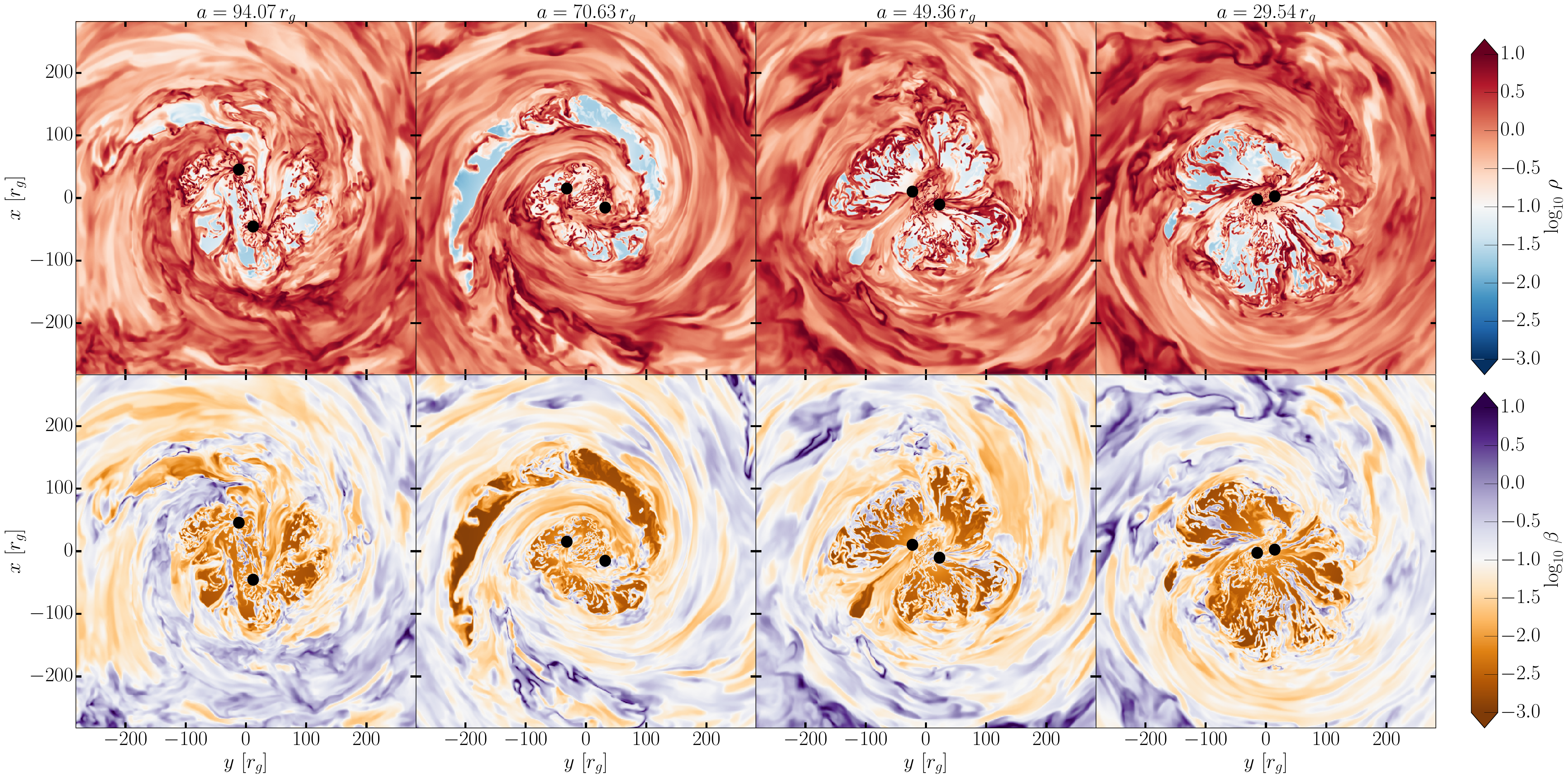}
    \caption{Inspiral of a supermassive black hole binary inside a magnetically arrested accretion disk starting from an initial separation $a_0 = 113.3 r_g$, where $r_g = GM /c^2$. From left to right, we show four different times throughout the inspiral, which are characterized by the orbital separation $a$.
    Initially (left panel), the binary is fully coupled to the accretion disk, and features enhanced flux ejection and Rayleigh-Taylor accretion streams characteristic for the magnetically arrested regime. During decoupling, large scale asymmetric streams appear as a result of mixing between two giant Rayleigh Taylor accretion streams (second panel). Only the outer most part of this magnetic flux tail will be ejected and mixed into the disk. Finally, the binary decouples (left and right panels), with the cavity wall remaining roughly at a constant location. Mass accretion onto the binary is still present in this regime, albeit primarily through Rayleigh-Taylor fingers. Instead of gradually spiraling inwards, the fingers will penetrate the cavity quasi-radially, free-falling onto the binary.
    Shown in color are the mass density, $\rho$, and plasma parameter, $\beta$.}
    \label{fig:inspiral}
\end{figure*}

\section{Introduction}\label{sec:intro}

Merging supermassive black hole (SMBH) binaries are prime targets for space-based gravitational wave detectors, such as LISA \cite{2017arXiv170200786A}, as well as for pulsar timing arrays, such as NANOGrav \cite{NANOGrav:2023gor}. 
Unlike stellar mass black hole binaries, for favorable mass ranges they are expected to remain within the detector band for weeks to years \cite{2017arXiv170200786A}, potentially enabling multi-messenger counterpart observations (especially of transient signals) of their late inspiral and merger (see, e.g., Refs. \cite{Bogdanovic:2021aav,LISA:2022yao} for recent reviews).

SMBH binaries are generally believed to be surrounded by a gaseous
environment (circumbinary disk) by the time of merger, due to the enhanced
gas inflow during their assembly, e.g., through galaxy collisions
\cite{Begelman1980,2006ApJS..163....1H,2007Sci...316.1874M,2008ApJS..175..356H}. 
At large binary separations, the gas disk is coupled to the binary through gravitational interactions (see, e.g., Ref. \cite{Lai:2022ylu} for a recent review). The gravitational tidal torque of the binary can clear up a cavity at the radial location where it balances the viscous torque of the disk \cite{Artymowicz:1994bw,Artymowicz:1996zz}. Accretion then proceeds largely in tidal accretion streams onto the binary masses \cite{Artymowicz:1994bw}.
Approaching merger, the orbital evolution of the binary is solely dominated by the emission of gravitational radiation \cite{Peters:1964zz}. 
As the binary spirals inwards, it can decouple from its host circumbinary disk when the binary inspiral speed surpasses the cavity closing speed \cite{2002ApJ...567L...9A,Dittmann2023a}.

To understand how and what type of transients should be expected (and ideally be characteristic to the presence of two black holes), it is imperative to understand the gas dynamics surrounding the black holes from the point of decoupling until right after merger \cite{2003MNRAS.340..411L,2005ApJ...622L..93M,2015MNRAS.447L..80F}. 
It has been suggested that the rapidly inwards spreading cavity wall is a potential source for X-ray emission \cite{2005ApJ...622L..93M,Westernacher-Schneider:2023cic,Krauth:2023nyq}. Notwithstanding accretion from mini-disks around each SMBH \cite{Bowen:2017oot,2022ApJ...928..187C,2021ApJ...910L..26P}, the complete absence of a large-scale accretion flow onto the SMBHs after decoupling could weaken potential jet signatures \cite{2003MNRAS.340..411L}. However, three-dimensional magnetohydrodynamical (MHD) simulations (see also Refs. \cite{2012ApJ...749..118S,Noble:2012xz,Noble:2021vfg,Most:2024qus}) indicate that the cavity might continue to shrink inwards \cite{Avara:2023ztw}, providing additional accretion onto the binary.

In a large number of works using tools from numerical relativity, it has been shown that the accretion of vertical flux from large scales is sufficient to power (dual) jets from the SMBHs, which cannot only serve as potential radio sources (see, e.g., Ref. \cite{Palenzuela:2009hx,Palenzuela:2010nf,Palenzuela:2010xn,Alic:2012df,Moesta:2011bn}), but reconnection in their interaction will be able to drive non-thermal emission in the X-ray (and potentially even gamma-ray) band \cite{Gutierrez:2023yor,Gutierrez:2021png,Ressler:2024tan,Krauth:2023nyq}. The merger itself may also be accompanied by copious flaring activity \cite{Ressler:2024tan}, akin to models proposed for neutron star binary precursor emission \cite{Most:2020ami,Most:2022ayk,Most:2022ojl,Most:2023unc}.

However, it remains an open question under which realistic conditions enough large vertical net flux could be provided during decoupling to power such jets. 
In particular, Refs. \cite{2011PhRvD..84b4024F,2014PhRvD..90j4030G} have investigated accretion from an artificial matter torus closely surrounding the binary over tens of orbits, whereas Ref. \cite{Avara:2023ztw} has used a relaxed configuration obtained from integrating the outer disk separately over hundreds of orbits, importing it as a boundary into their simulation. However, in those cases the magnetic fields present in the accretion flows were weak, making the overall flow resemble a standard and normal accretion (SANE) scenario onto a single black hole.

Motivated by recent findings that accretion disks formed in cosmological simulations may be strongly magnetized \cite{Shi2024}, we have demonstrated in an earlier work \cite{Most:2024qus} that the outer accretion flow can become magnetically arrested (MAD) \cite{Narayan2003,Igumenshchev:2007bh,Tchekhovskoy:2011zx}. In this case, the cavity is filled with strong vertical magnetic flux, naturally providing conditions for dual jets to be launched \cite{Most:2024qus}.

In this work, we follow the evolution of such a magnetically arrested circumbinary disk through decoupling and merger, and show that enhanced flux ejections from the cavity will set in during decoupling and shortly after merger.

\section{Methods}

In this work, we numerically integrate the evolution of an equal-mass, circular binary inside a magnetically arrested circumbinary disk \cite{Most:2024qus}. 
The masses are modeled as two Newtonian point masses, $M_1 = M_2$, with orbital separation, $a$. 
The resulting gravitational potential $\Phi$ is smoothened near those masses using a Plummer prescription \cite{1911MNRAS..71..460P}. On top of this background, we solve the Newtonian ideal MHD equations in flux conservative form on a Cartesian grid using the   \texttt{AthenaK} \cite{AthenaK} performance portable version of the \texttt{Athena++} code \cite{2020ApJS..249....4S}, which utilizes the \texttt{Kokkos} library\cite{Kokkos}. We further adopt a locally isothermal equation of state, where the pressure $P=\rho c_s^2(r)$, is calculated based on the local mass density, $\rho$, and the sound speed, $c_s = \sqrt{\Phi/\mathcal{M}}$, which is fixed by the gravitational potential, and the sonic Mach number, $\mathcal{M}=10$ (namely, the disk aspect ratio is $h=1/\mathcal{M}=0.1$). 
The sink prescription near the point masses is given by a locally Keplerian mass removal scheme with a sink radius of $r_{\rm sink}= 0.07a$ using the same prescription as Ref. \cite{Most:2024qus}. We further adopt a floor for the plasma parameter, $\beta = 2P/B^2$, where $B$ is the magnetic field strength, such that $\beta > 10^{-3}$. We also adopt a first-order flux correction scheme to correct problematic cells \cite{Lemaster:2008gh}. For the stability of the sink region, within half of the sink radius (depletion radius $r_{\rm dep}=0.5r_{\rm sink}$), we fix the mass density, pressure, and plasma parameter to be the floor value, the velocity is fixed to match that of the binary.\\
Since we are mainly interested in how the binary inspirals and decouples from the disk due to the emission of gravitational waves, we use our previously computed circumbinary disk state after about $150$ orbits as the starting point for this work \cite{Most:2024qus}. At this time, the central cavity region is strongly magnetized, $\beta \simeq 10^{-3}$, and accretion has entered a quasi-periodic MAD accretion cycle.\\
We then evolve the orbit according to lowest order gravitational wave radiation reaction \cite{Peters:1964zz}.
That is, we adopt a physical initial separation of $a_0 = 113.3 r_g$, where $r_g = G M /c^2$ is the gravitational radius, and then impose the inspiral according to
\cite{1964PhRv..136.1224P},
    $a\left(t\right) = a_0 \left[1- {t}/{\tau_{\rm GW}}\right]^{1/4}$,
where the characteristic merger time is given as 
    $\tau_{\rm GW} = 5 a_0^4/\left[256 M_1^3 q \left(q + 1\right)\right]$,
where $q = M_2/M_1$ is the mass ratio.
We adopt the same computational domain as Ref. \cite{Most:2024qus}: An outer boundary of $\pm 80\, a_0$, six levels of mesh refinement, and a finest resolution of $\Delta x = 0.0065 a_0$. The simulations have been carried out on 600 GPUs on the OLCF Summit system using about $60,000$ V100-GPU hours.

\begin{figure*}
    \centering
    \includegraphics[width=\linewidth]{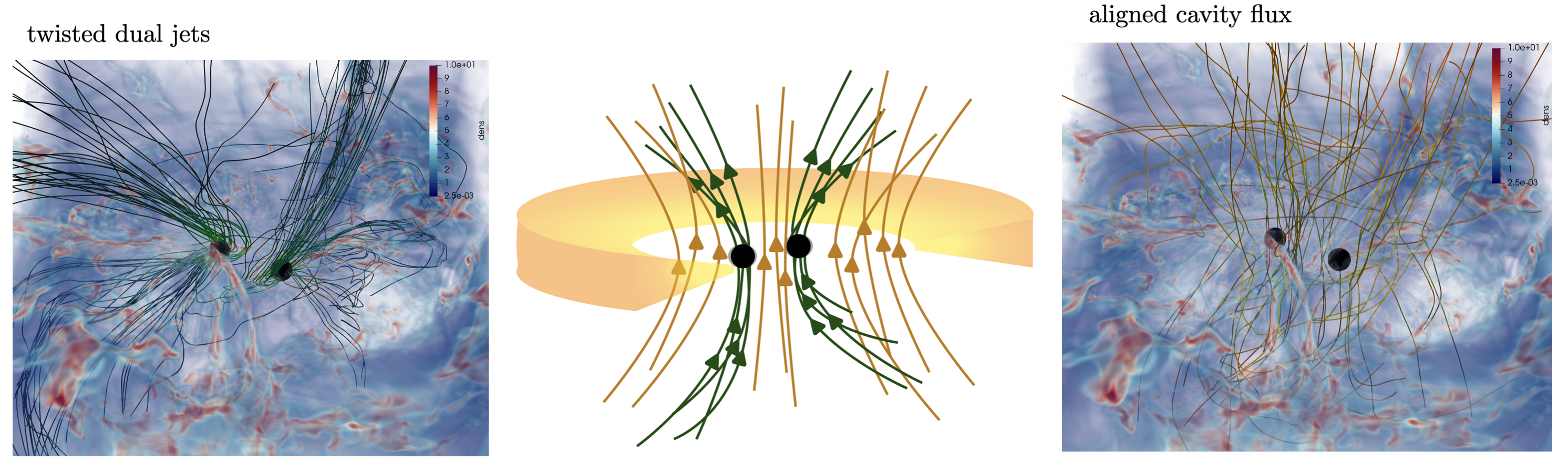}
    \caption{Magnetic field topology inside the cavity during decoupling for separation of $a = 50\, r_g$. The magnetic field can be decomposed into different segments. \textit{(Left)} Green field lines denote magnetic flux tubes threading the black holes. These are dragged along in a geometry resembling a dual jet configuration. \textit{(Right)} Yellow field lines indicate the strong large-scale net vertical flux present in the cavity after decoupling.}
    \label{fig:3d-cavity}
\end{figure*}

\section{Results}

In this work, we investigate the decoupling of a SMBH binary from its magnetically arrested circumbinary disk.
 
Starting out in a magnetically arrested state \cite{Most:2024qus}, the circumbinary disk cavity is saturated with strong vertical magnetic field flux. Even in the quasi-steady state of accretion, the cavity edge is interchange unstable, the accretion onto the binary proceeds through material spirals connecting to the disk and Rayleigh-Taylor accretion streams \cite{Spruit:1995fr,Kulkarni:2008vk}. An unstable cavity also leads to magnetic flux eruptions from the cavity \cite{Tchekhovskoy:2011zx,Ripperda:2021zpn,Chatterjee2022,Begelman:2021ufo}.

\subsection{Inspiral}
Different from a binary on a constant orbit, a shrinking binary reduces the gravitational torque on the disk, causing the disk cavity wall to move inwards \cite{2003MNRAS.340..411L,2015MNRAS.452.2396M,2015MNRAS.447L..80F,Dittmann2023a}.
In the magnetically arrested state, this inward motion is accompanied by an even more interchange unstable \cite{Spruit:1995fr,Begelman:2021ufo} cavity wall, causing larger and more frequent flux eruptions, which then in the isothermal (instantaneous cooling) approximation we use, get rapidly sheared and mixed into the disk (see left panel Fig. \ref{fig:inspiral}). 
As the binary continues to spiral inward, at some point the decoupling radius will be reached (when the shrinking speed of the binary's semi-major axis becomes faster than the closing speed of the cavity wall \cite{2002ApJ...567L...9A,Dittmann2023a}). 
Around this time (when $a\sim 70\, r_g$), the cavity becomes strongly lopsided and two large Rayleigh Taylor accretion streams merge into a prominent one-armed spiral structure in the cavity (see second panel from the left in Fig. \ref{fig:inspiral}). 
Similar features of giant stream mixing and formation of spiral-shaped flux tube can also be observed even for a quasi-steady state of magnetically arrested disk \cite{Wang2024}, however, the inspiral seems to significantly enhance this feature. 
Similar to flux eruptions shown at earlier times (Fig. \ref{fig:inspiral}, left panel), the actual ejected flux tubes represent only a small fraction of the cavity size. The precise dynamics and subsequent propagation will depend strongly on the choice of disk thermodynamics (cooling and equation of state) \cite{Wang2024}, see also Ref. \cite{Vos:2023day}.
Nonetheless, this enhanced lopsided cavity appears to be characteristic of the near-decoupling regime. \\
We also observe that this large coherent flux tube is threaded by secondary Rayleigh-Taylor fingers. 
These have been shown to be perfect sites for particle acceleration \cite{Zhdankin:2023wch}, potentially powering flares \cite{Porth:2020txf,Dexter:2020cuv,Ripperda:2021zpn}.
We will quantify this further in Sec. \ref{sec:emission}.

\subsection{Decoupling}
\begin{figure}
    \centering
    \includegraphics[width=\linewidth]{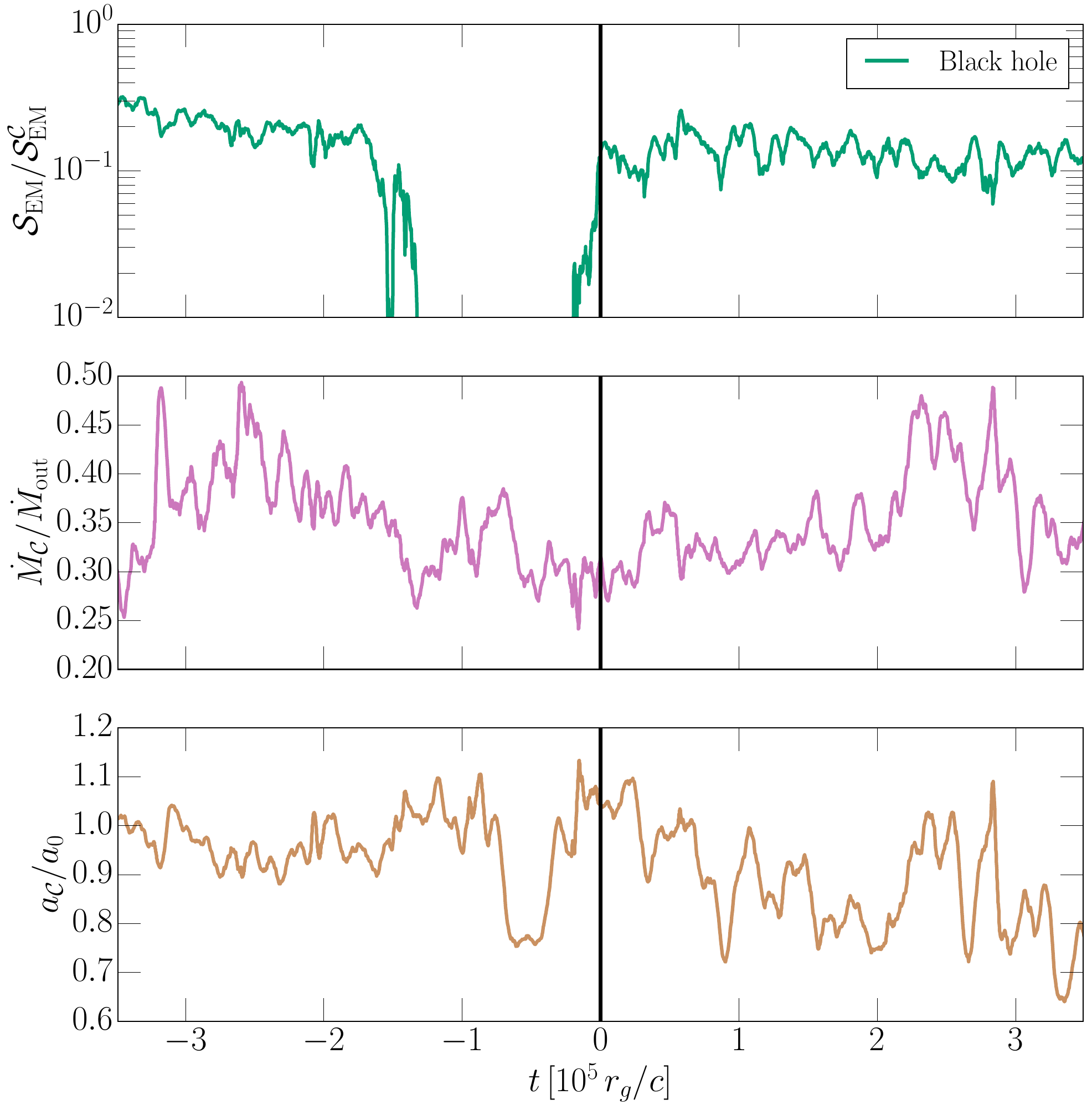}
    \caption{Accretion diagnostics: {\it (Top)} Electromagnetic energy (Poynting) flux, $\mathcal{S}_{\rm EM}$, of the black hole normalized to the total Poynting flux of the cavity, $\mathcal{S}_{\rm EM}^\mathcal{C}$; 
    {\it (Center)} Mass accretion rate, $\dot{M}_\mathcal{C}$, into the cavity normalized by the instantaneous average outer accretion rate, $\dot{M}_{\rm out}$; {\it (Bottom)} Cavity radius, $a_{\mathcal{C}}$, relative to the initial binary separation, $a_0$. The time of merger is marked by a black vertical line, and times are stated relative to the gravitational timescale $r_g/c = GM/c^3$.}
    \label{fig:poynting}
\end{figure}

At separation around $a\simeq 50 r_g/c$, we observe a constant shape of the cavity (compare Fig. \ref{fig:inspiral}, right panels), indicating the binary has decoupled.
Decoupling approximately happens when the radial acceleration of the ingoing gas is smaller than that of the outer (Keplerian) accretion flow, leading to an approximate decoupling radius of \cite{2002ApJ...567L...9A,Dittmann2023a},
$ {a_{\rm decouple}}/{a_0} = \sqrt{\xi {\nu_{\rm GW}}/{\nu_{\rm hydro}}}$,
where $\xi \simeq 0.3-0.5$ \cite{Dittmann2023a}, and $\nu_{\rm hydro}$ and $\nu_{\rm GW} = a_0^2/\tau_{\rm GW} $ are the effective viscosity of the accretion flow and gravitational wave emission. In our case, $\nu_{\rm hydro}$ is self-consistently provided by the MHD flow. In order to quantify the impact of decoupling on the accretion flow, we compute the effective cavity radius, $a_\mathcal{C}$, as the radial location where the azimuthally and vertically averaged disk density drops by a factor ten (Fig. \ref{fig:poynting}, bottom panel). We can see that the cavity size does not appreciable change, consistent with low viscosity cases reported by Ref. \cite{Dittmann2023a}.\\
At this stage, previous work has speculated that accretion might cease \cite{2003MNRAS.340..411L,2005ApJ...622L..93M}, with numerical simulations indicating continued but reduced accretion rates, in part driven by eccentric dynamics of the cavity (e.g., \cite{Avara:2023ztw}). 
For a magnetically arrested flow, we find another natural way of continued mass transport:
even though the gravitational torque at the disk cavity wall is significantly reduced, the magnetic pressure in the cavity can temporarily balance the effective viscous torque in the disk.
This implies, that the cavity wall will continue to be (even more) interchange unstable \cite{Spruit:1995fr,Kulkarni:2008vk,Parfrey:2023swe,Murguia-Berthier:2023fji}. Indeed, we observe the presence of Rayleigh-Taylor fingers in the decoupled cavity at virtually all times.  This feature, which is unique to magnetospheric accretion \cite{Kulkarni:2008vk,Parfrey:2023swe}, does not seem to affect the overall magnetization of the cavity with $\beta \gtrsim 10^{-3}$.

Since these final stages of the inspiral process can directly be compared to full general-relativistic(GR) MHD simulations covering separations up to $25\, r_g$ (e.g., \cite{Ressler:2024tan}), 
we now separate the magnetic field components in the cavity into two parts, based on whether these magnetic fields are directly connected to the black holes (sinks) or are only contained in the cavity (Fig. \ref{fig:3d-cavity}). 
We can see the magnetic field topology inside the decoupled cavity has two characteristic features: Overall, the cavity is filled with strong vertical net flux (yellow). This flux has accumulated in the cavity over many accretion cycles prior to decoupling \cite{Most:2024qus}, and is strongly affected by large-scale flux supply owing to the formation of the circumbinary disk \cite{2023arXiv231004506H,Shi:2024skj}, as well as to the thermodynamics of the disk \cite{Wang2024}. It is the presence of this net flux going all the way from the binary to the cavity wall, which enables  stream-like magnetospheric accretion onto the binary after decoupling.

\begin{figure*}
    \centering    
    \includegraphics[width=\linewidth]{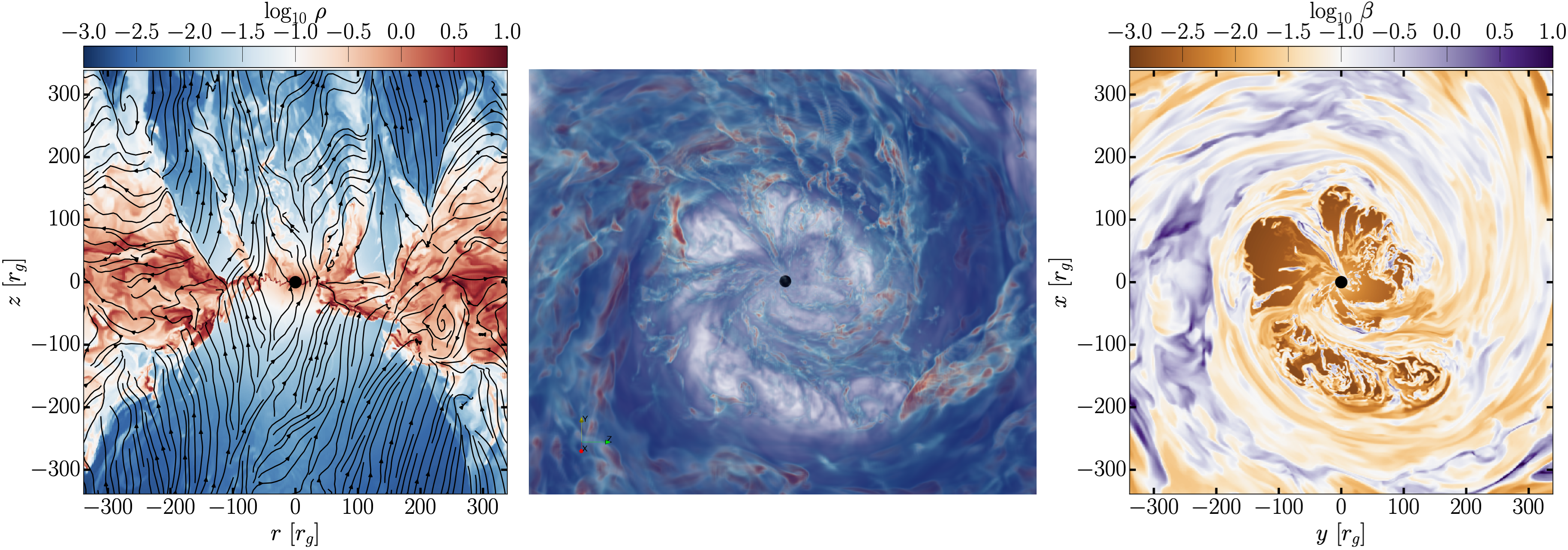}
    \caption{Post-merger accretion flow at $t=9.3\times10^4 r_g/c$. {\it (Left)} Mass density, $\rho$, in the meridional plane. Magnetic field lines are denoted by black streamlines, exhibiting a magnetic tower structure indicative of jet formation. Erupting magnetic flux is shown in terms of {\it (Center)} mass density, and {\it (Right)} magnetization parameter, $\beta$, in the equatorial plane. }
    \label{fig:merger-comp}
\end{figure*}
Overall, the condition of the binary embedded in a strong vertical net flux matches (without any disk) many of the initial conditions adopted in three-dimensional numerical relativity work of merging black hole binaries (e.g., \cite{Giacomazzo:2012iv,Kelly:2017xck,Kelly:2020vpv,Cattorini:2021elw,Cattorini:2021elw}).
Although largely ad hoc in those studies, our results provide a natural formation channel for this merger scenario.
As a consequence, the accretion of large-scale flux onto the sink then enables magnetic tower driven outflows \cite{Lynden-Bell:2002voo}, resembling jets (green field lines in Fig. \ref{fig:3d-cavity}). These outflows are highly bent owing to the motion of the sinks. Similar phenomenology can also be seen in magnetized accretion onto moving black holes with the degree of bending corresponding to the effective wind speed \cite{Kim:2024zjb,Kaaz:2022dsa,Cayuso:2019uok}. We stress that even for the accretion of net toroidal flux from the disk jet-like outflows can be powered \cite{Kim:2024zjb}, which in our case will be enabled by buoyant instabilities of the toroidal magnetic field around the sink (see, e.g., Refs. \cite{Most:2023sft,Musolino:2024sju} for MHD simulations and discussions of related phenomena in neutron star mergers). As a result the outflows are likely weak, and in our case intermittent as they require continuous feeding of the sink through accretion streams. This scenario is not unlike the weak and intermittent jet launching case reported in Ref. \cite{Ressler:2024tan}, but it may change depending on the (potential) presence of mini-disks around the black holes \cite{Bowen:2017oot,Combi:2021dks,Paschalidis:2021ntt}, which could continue to support a MAD state of each SMBH.\\

We quantify the amount of jet-like outflows via the outgoing Poynting flux, $\mathcal{S}_{\rm EM}$ on each sink (Fig. \ref{fig:poynting}, top panel). We can now clearly see that the outflow power steadily decreases, until it completely ceases
abruptly around $1.5\times 10^5\, r_g/c$ before merger. This indicates that mass accretion onto the sinks has become insufficient to power strong magnetically driven outflows. At the same time the mass accretion rate, $\dot{M}_{\mathcal{C}}$, into the cavity (Fig. \ref{fig:poynting}, center panel) has only moderately decreased. We therefore find it useful to define decoupling via the lack of outflows from the sink (however, see also Ref. \cite{Dittmann2023a}).

\subsection{Merger}

Finally, the emission of gravitational waves will drive the binary to collision. This process has been extensively studied for very small separations using GRMHD simulations in dynamical spacetimes \cite{Farris:2012ux,2014PhRvD..89f4060G,2014PhRvD..90j4030G,Paschalidis:2021ntt,Bright:2022hnl,2017ApJ...838...42B,2021ApJ...913...16L,Avara:2023ztw,Ressler:2024tan}. Notable features of the merger include, pre-merger flaring \cite{Ressler:2024tan}, jet interaction \cite{Palenzuela:2010nf,Gutierrez:2023yor,Ressler:2024tan}, mini-disk dynamics \cite{2017ApJ...838...42B,2021ApJ...913...16L,2022ApJ...928..187C,Paschalidis:2021ntt}. Here, we focus on the novel case where the SMBHs merge in the magnetically arrested cavity with extremely strong vertical net flux, not considered before. 
While ideally a full GR treatment would be desirable to capture feedback from the SMBH(s), we here mainly focus on the cavity (wall) dynamics and associated transients during and after merger, which remains located at $100 r_g$. For a first exploration of these aspects, a Newtonian MHD treatment enabling a self-consistent treatment of the outer accretion flow is likely sufficient.

Since the black holes have rapidly merged inside the cavity, there is a sudden loss of the supporting gravitational torque on the outer accretion flow. In other words, the cavity wall is now purely supported hydro-magnetically.
Once interchange instabilities at the cavity wall develop, this will lead to a large inflow of matter onto the merged black hole, with a subsequent ejection of vertical flux. This can be seen in Fig. \ref{fig:merger-comp}. More specifically, we see the development of a large scale one-arm accretion stream into the cavity, with the subsequent ejection of a large evacuated (magnetically dominated) region, roughly spanning up to a third of the cavity volume.
These large-scale flux eruptions are larger and more violent than regular horizon-scale eruptions around isolated black holes \cite{Tchekhovskoy:2011zx,Ripperda:2021zpn,Vos:2023day}.
In particular, we can see large Rayleigh-Taylor unstable regions inside the ejected flux tube, providing net acceleration sites for non-thermal emission \cite{Zhdankin:2023wch}. In addition to quasi-periodicity in the flux eruptions leading up to and during decoupling, this feature may be able to power a flare-type transient \cite{Dexter:2020cuv,Porth:2020txf,Ripperda:2021zpn}, specific to the magnetically arrested case we discuss.\\

\subsubsection{Emission signatures}\label{sec:emission}
We will now provide some simple estimates for the expected electromagnetic emission.
Within our simulation using an isothermal cooling approximation (which in essence couples temperature to gravitational distance), we can assign an effective plasma parameter $\sigma$, via
\begin{align}\label{eqn:sigma}
    \sigma \simeq \frac{B^2}{\rho c^2} = \frac{2}{\mathcal{M} \beta} \frac{r_g}{r} = 200 \, \left(\frac{10^{-3}}{\beta}\right) \left(\frac{10}{\mathcal{M}} \right) \frac{r_g}{r}\,,
\end{align}
where r the distance to the remnant black hole. Near the cavity wall, where reconnection occurs inside ejected flux tubes, we then generally have $\sigma <10$. This is a natural consequence of the outer parts of the cavity transitioning to the disk and should be valid for all flux eruptions we see.
Reconnection can now happen in two places in such a regime, akin to isolated black hole magnetospheres \cite{Ripperda:2020bpz,Ripperda:2021zpn}. Close to the black hole, reconnection will happen in a radiative regime \cite{Ripperda:2021zpn}. There, synchrotron losses balance acceleration by the electric field of the layer \cite{Uzdensky:2011df}, leading to an approximately constant Lorentz factor, $\gamma_{\rm rad} \simeq \sqrt{3 c \beta_{\rm rec} / 2 r_e \omega_B} $, where $\beta_{\rm rec} \simeq 0.1$ \cite{Sironi:2014jfa} is the reconnection rate, $r_e$ the classical electron radius, and $\omega_B = e B /\left( m_e c\right)$ the cyclotron frequency, with $e$ and $m_e$ being the electron charge and mass, respectively. Since the mass density inside the cavity is approximately constant (see Fig. \ref{fig:inspiral}), we assume based on Eq. \eqref{eqn:sigma}, that $B(r) = B_H \sqrt{r_g/r}$, where $B_H$ is the magnetic field strength near the horizon.
However, in the low-$\sigma$ regions near the cavity wall we may not be in such a regime \cite{Ripperda:2021zpn,Zhdankin:2023wch}, meaning that the particle Lorentz factor reaches at most $\gamma_{\rm sync} \simeq \sigma_e$, where $\sigma_e = \sigma m_p/m_e$ is the electron plasma parameter, and $m_p$ the proton mass. We can then estimate the average photon energy via 
\begin{align}
E_{\gamma} &= \gamma_{\rm sync}^2 \hbar \omega_B \\
           &\simeq 1.6\, eV\, \left(\frac{B_H}{100\,\rm G}\right)\left(\frac{100\, r_g}{r}\right)^{1/2}\left(\frac{\sigma}{2}\right)^2.
\end{align}
This conservatively places flares from flux eruptions into the near infrared band, though depending on the exact location and magnetization of emitting region, the emission will likely be multi-band. 
The timescale of the flares after merger, is about $t_{\rm flare} \simeq 9 \times 10^4 r_g/c \simeq 5 \left(M/10^6\, M_\odot\right)\, \rm days$.
Emission from these flares would be in addition to, e.g., vanishing of thermal X-ray after decoupling \cite{Krauth:2023nyq}, appearance of non-thermal X-ray from jet interactions \cite{Gutierrez:2023yor,Ressler:2024tan}, or pre-merger flares \cite{Ressler:2024tan}.

\section{Conclusions}
In this work, we have provided a first investigation of a SMBH binary decoupling from its magnetically arrested accretion environment. 
Starting from our previously computed quasi-steady state \cite{Most:2024qus}, we have simulated the fast inspiral and decoupling of the SMBH binary solely driven by gravitational wave emission.

During the inspiral, the shrinking of the binary's separation leads to a reduction of disk resonances at all radii. As a result, the strength of the gravitational torque at the cavity wall reduces continuously. The cavity wall then becomes unstable to interchange instabilities even before it moves inward to re-establish a new equilibrium. 
In turn, magnetic flux eruptions near the cavity wall are enhanced near decoupling and merger.
These have potential non-thermal signatures \cite{Ripperda:2021zpn,Hakobyan:2022alv,Porth:2020txf} (e.g., through internal Rayleigh-Taylor instabilities \cite{Zhdankin:2023wch}), and also modify the angular momentum budget of the accretion disk \cite{Chatterjee2022}. 
As the binary orbit shrinks, the system fully decouples and the binary becomes embedded in a hyper-magnetized cavity. We identify this point with the absence of accretion-powered electromagnetic outflows (Fig. \ref{fig:poynting}), which is overall consistent with the picture of vanishing mini-disks, thought to lead to a fading of thermal X-ray emission prior to merger \cite{Krauth:2023nyq}. 
That said, we find that in our cavity with strong vertical net flux, magnetic-tower-like outflows (dual jets) can be launched. 
In a full GRMHD sense, the individual BHs can power outflows by accreting this net vertical field readily present in the cavity. 
As such, a magnetically arrested cavity provides the ideal setting for dual jet formation \cite{Palenzuela:2009hx,Palenzuela:2010nf,Moesta:2011bn,Farris:2012ux,2014PhRvD..89f4060G,2014PhRvD..90j4030G}, with a potential appearance of non-thermal X-ray emission from jet interactions \cite{Ressler:2024tan,Gutierrez:2023yor}. 

Going through merger, we find there is an immediate relaunching of electromagnetic outflows at and after merger, 
happening on much faster time scales than the cavity closing time due to viscous relaxation\cite{2005ApJ...622L..93M}.
This is because the magnetically arrested cavity wall (in the absence of a gravitational torque from the binary) becomes immediately interchange unstable \cite{Spruit:1995fr}, with Rayleigh-Taylor streams providing accretion onto the newly merged black hole on the effective free-fall timescale. 
Finally, after merger we find that the cavity is unstable to large-scale flux eruptions. This is because the large magnetically supported cavity is likely not a stable force-free configuration that will ultimately need to relax to a MAD accretion flow.
During these eruptions, we expect substantial non-thermal emission to be present \cite{Ripperda:2021zpn,Zhdankin:2023wch},
which will likely be multiband, and in particular near-infrared on timescales of days to months after merger, depending on the binary mass.
Systematic follow-up work should be devoted toward clarifying these emission signatures.\\
However, our Newtonian simulations do not fully allow us to properly quantify the binary merger process and this post-merger dynamics. Meanwhile, the equation of state of the accretion flow can also significantly alter the dynamics of decoupling and post-merger magnetic flux ejection.
Instead, our work highlights the promise and importance of investigating the magnetically arrested regime surrounding SMBH binaries for hundreds of orbits using fully dynamical-spacetime GRMHD simulations (potentially with added microphysics), which have recently become feasible \cite{Ressler:2024tan,Ressler:2024mpx,Combi:2024inn}.\\

\begin{acknowledgments}
The authors are grateful for insightful discussions with Xue-Ning Bai, Luciano Combi, Philip F. Hopkins, Amir Levinson, Douglas N. C. Lin, Sean Ressler, Bart Ripperda, James M. Stone, and Alexander Tchekhovskoy. 
The simulations were performed on DOE OLCF Summit under allocation AST198. 
This research used resources from the Oak Ridge Leadership Computing Facility at the Oak Ridge National Laboratory, which is supported by the Office of Science of the U.S. Department of Energy under Contract No. DE-AC05-00OR22725.
Additional simulations were done on DOE NERSC supercomputer Perlmutter under grant m4575.
This research used resources of the National Energy Research
Scientific Computing Center, a DOE Office of Science User Facility
supported by the Office of Science of the U.S. Department of Energy
under Contract No. DE-AC02-05CH11231 using NERSC award
NP-ERCAP0028480.
\end{acknowledgments}

\newcommand\aj{\refjnl{AJ}}%
\newcommand\psj{\refjnl{PSJ}}%
\newcommand\araa{\refjnl{ARA\&A}}%
\renewcommand\apj{\refjnl{ApJ}}%
\newcommand\apjl{\refjnl{ApJL}}     %
\newcommand\apjs{\refjnl{ApJS}}%
\renewcommand\ao{\refjnl{ApOpt}}%
\newcommand\apss{\refjnl{Ap\&SS}}%
\newcommand\aap{\refjnl{A\&A}}%
\newcommand\aapr{\refjnl{A\&A~Rv}}%
\newcommand\aaps{\refjnl{A\&AS}}%
\newcommand\azh{\refjnl{AZh}}%
\newcommand\baas{\refjnl{BAAS}}%
\newcommand\icarus{\refjnl{Icarus}}%
\newcommand\jaavso{\refjnl{JAAVSO}}  %
\newcommand\jrasc{\refjnl{JRASC}}%
\newcommand\memras{\refjnl{MmRAS}}%
\newcommand\mnras{\refjnl{MNRAS}}%
\renewcommand\pra{\refjnl{PhRvA}}%
\renewcommand\prb{\refjnl{PhRvB}}%
\renewcommand\prc{\refjnl{PhRvC}}%
\renewcommand\prd{\refjnl{PhRvD}}%
\renewcommand\pre{\refjnl{PhRvE}}%
\renewcommand\prl{\refjnl{PhRvL}}%
\newcommand\pasp{\refjnl{PASP}}%
\newcommand\pasj{\refjnl{PASJ}}%
\newcommand\qjras{\refjnl{QJRAS}}%
\newcommand\skytel{\refjnl{S\&T}}%
\newcommand\solphys{\refjnl{SoPh}}%
\newcommand\sovast{\refjnl{Soviet~Ast.}}%
\newcommand\ssr{\refjnl{SSRv}}%
\newcommand\zap{\refjnl{ZA}}%
\renewcommand\nat{\refjnl{Nature}}%
\newcommand\iaucirc{\refjnl{IAUC}}%
\newcommand\aplett{\refjnl{Astrophys.~Lett.}}%
\newcommand\apspr{\refjnl{Astrophys.~Space~Phys.~Res.}}%
\newcommand\bain{\refjnl{BAN}}%
\newcommand\fcp{\refjnl{FCPh}}%
\newcommand\gca{\refjnl{GeoCoA}}%
\newcommand\grl{\refjnl{Geophys.~Res.~Lett.}}%
\renewcommand\jcp{\refjnl{JChPh}}%
\newcommand\jgr{\refjnl{J.~Geophys.~Res.}}%
\newcommand\jqsrt{\refjnl{JQSRT}}%
\newcommand\memsai{\refjnl{MmSAI}}%
\newcommand\nphysa{\refjnl{NuPhA}}%
\newcommand\physrep{\refjnl{PhR}}%
\newcommand\physscr{\refjnl{PhyS}}%
\newcommand\planss{\refjnl{Planet.~Space~Sci.}}%
\newcommand\procspie{\refjnl{Proc.~SPIE}}%

\newcommand\actaa{\refjnl{AcA}}%
\newcommand\caa{\refjnl{ChA\&A}}%
\newcommand\cjaa{\refjnl{ChJA\&A}}%
\newcommand\jcap{\refjnl{JCAP}}%
\newcommand\na{\refjnl{NewA}}%
\newcommand\nar{\refjnl{NewAR}}%
\newcommand\pasa{\refjnl{PASA}}%
\newcommand\rmxaa{\refjnl{RMxAA}}%

\bibliography{inspire,non_inspire,cbd}%

\end{document}